\documentclass[pra,fleqn,showpacs,twocolumn]{revtex4}

\usepackage{amsmath,graphicx}

\begin{document}

\title{Calculation of transition probabilities and ac Stark shifts in two-photon laser transitions of antiprotonic helium}
\author{Masaki Hori}
\affiliation{Max-Planck-Institut f\"ur Quantenoptik, Hans-Kopfermann-Strasse 1, 85748 Garching, Germany}
\affiliation{Department of Physics, University of Tokyo, Hongo, Bunkyo-ku, Tokyo 113-0033, Japan}
\author{Vladimir I.~Korobov}
\affiliation{Joint Institute for Nuclear Research 141980, Dubna, Russia}

\begin{abstract}
Numerical {\em ab initio} variational calculations of the transition probabilities and ac Stark shifts in two-photon transitions
of antiprotonic helium atoms driven by two counter-propagating laser beams are presented. We found that sub-Doppler
spectroscopy is in principle possible by exciting transitions of the type $(n,L)$$\rightarrow$$(n-2,L-2)$ between antiprotonic
states of principal and angular momentum quantum numbers $n\sim L-1\sim 35$, first by using highly monochromatic, nanosecond laser beams of
intensities $10^4-10^5$ W/cm$^2$, and then by tuning the virtual intermediate state close (e.g., within 10--20 GHz) to the real state $(n-1,L-1)$ to enhance
the nonlinear transition probability. We expect that ac Stark shifts of a few MHz or more will become an important source of systematic error
at fractional precisions of better than a few parts in $10^9$. These shifts can in principle be minimized and even canceled
by selecting an optimum combination of laser intensities and frequencies. We simulated the resonance profiles of some
two-photon transitions in the regions $n=30$--40 of the $\overline{p}{\rm ^4He}^+$ and $\overline{p}{\rm ^3He}^+$ isotopes
to find the best conditions that would allow this.
\end{abstract}

\maketitle

\section{Introduction}

The transition frequencies $\nu_{\rm exp}$ of antiprotonic helium atoms \cite{Hayanorpp,PhysRep,Mor94,Maas,Tor99}
($\overline{p}{\rm He}^+\equiv \overline{p}^-+e^-+{\rm He}^{2+}$) have recently been measured by single-photon laser spectroscopy to
a fractional precision of $\sim 1$ part in $10^{8}$ \cite{Hori01,Hori03,Hori06}. By comparing these results with three-body QED
calculations \cite{Kor00,Kor03,Kor08,kino2004,andersson}, the antiproton-to-electron mass ratio has been determined as 1836.152674(5) \cite{Hori06,CODATA06}.
To further increase the experimental precision on $\nu_{\rm exp}$, we have proposed future experiments \cite{hori2000}
of sub-Doppler two-photon spectroscopy of $\overline{p}{\rm He}^+$ by irradiating the atom with two counter-propagating
laser beams \cite{horiopt09}. Dynamic (ac) Stark effects are expected to become one of the important sources of systematic error in
these future experiments, as is the case with other high-precision laser spectroscopy measurements of atomic hydrogen
\cite{Hansch86,garreau,fischer, Haas2006,Haas2006_2,kolachevsky} and antihydrogen \cite{gabrielse,andresen}, molecular hydrogen \cite{hannemann,hilico},
helium \cite{eikema1996,eikema1997,minardi,bergeson}, and muonium
\cite{yakhontov96,yakhontov99,meyer}. In this paper we calculate the transition probability and ac Stark shift involved in these
two-photon transitions using precise three-body wavefunctions of $\overline{p}{\rm He}^+$.

\begin{figure}
\begin{center}
\vspace*{6mm}
\hspace{-5mm}\includegraphics[width=66mm]{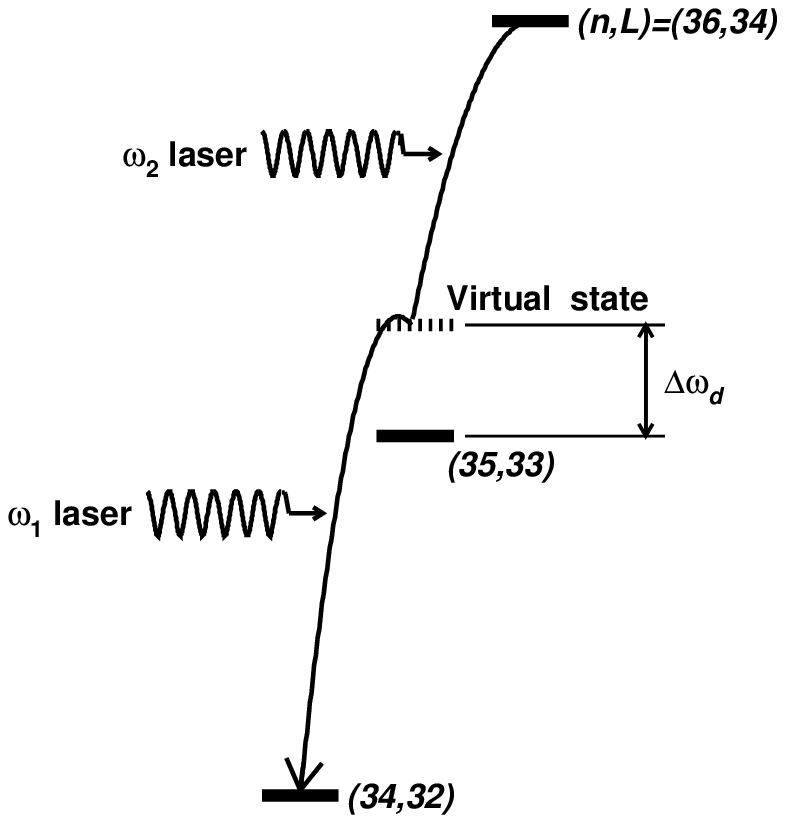}
\vspace{-7mm}
\end{center}
\caption{\label{energydig} Energy level diagram indicating the two-photon transition
$(n,L)=(36,34)$$\rightarrow$$(34,32)$ in $\overline{p}{\rm ^4He}^+$. The relative
position of the virtual intermediate state for a detuning frequency $\Delta\omega_d$
of the two counter-propagating laser beams is shown.}
\vspace{-3mm}
\end{figure}

\begin{figure*}[htbp]
\begin{center}
\vspace{-3mm}
\includegraphics[width=160mm]{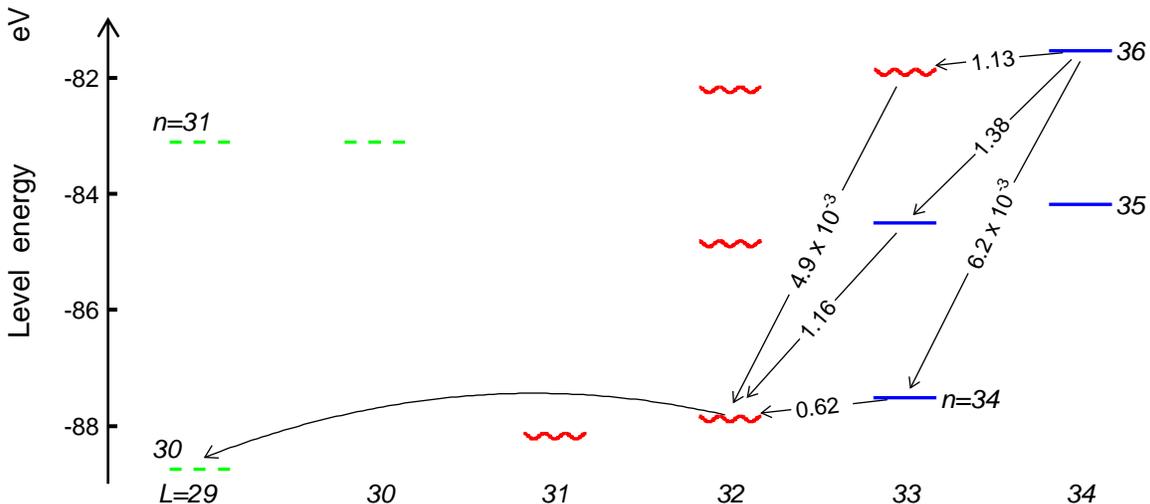}
\vspace{-3mm}
\end{center}
\caption{\label{energydig2} (Color online) portion of the energy level diagram of $\overline{p}{\rm ^4He}^+$.
The solid lines indicate radiation-dominated metastable states, the wavy lines Auger-dominated
short-lived states. The broken lines indicate $\overline{p}{\rm He}^{2+}$ ionic states formed
after Auger emission, the curved arrows Auger transitions with minimum
$|\Delta L|$. Calculated values of the dipole moments for some single-photon radiative
transitions are shown in atomic units.}
\end{figure*}

The $\overline{p}{\rm He^+}$ atoms \cite{PhysRep} can be easily synthesized by simply allowing antiprotons
to slow down \cite{knudsen,foster,luhr2009,henkel2009,mcgovern,barna} and come to rest in a helium target. Some of
the antiprotons are captured \cite{mhori2002,mhori2004,briggs1999,cohen2004,hesse,tokesi,ovchinnikov,revai06,tong08,genkin}
into Rydberg $\overline{p}{\rm He}^+$ states with large principal ($n\sim 38$) and angular momentum ($L\sim n-1$) quantum
numbers that have microsecond-scale lifetimes against antiproton annihilation in the helium nucleus.
The longevity is due to the ground-state electron in $\overline{p}{\rm He}^+$ which protects the antiproton during collisions with other helium atoms
\cite{mhori1997,russell02,obreshkov}. All laser spectroscopy experiments \cite{Hori06} reported so far have used
pulsed lasers \cite{ol2003} to induce
single-photon transitions of antiprotons occupying these metastable states, to short-lived states with nanosecond-scale
lifetimes against Auger emission of the electron \cite{yamaguchi2002,yamaguchi2004,kor97,revai97,kartavtsev2000}.
A Rydberg $\overline{p}{\rm He}^{2+}$ ion \cite{mhori2005,sakimoto2007,korenman07,sakimoto2009} then remained
after Auger decay, which was rapidly destroyed by collisional Stark effects.
The resulting resonance profiles of $\overline{p}{\rm He}^+$ had Doppler widths
$\Delta\omega_D/2\pi\sim 0.3-1.2$ GHz corresponding to the thermal motion of $\overline{p}{\rm He}^+$
in the experimental target at $T\sim10$ K. This broadening limited the experimental precision on $\nu_{\rm exp}$.

The first-order Doppler broadening can in principle be reduced (Fig.~\ref{energydig}) \cite{hori2000} by
irradiating $\overline{p}\mbox{He}^+$ atoms with two counter-propagating laser
beams of angular frequencies $\omega_1$ and $\omega_2$ and inducing, e.g., the
two-photon transition $(n,L)\!\to\!(n\!-\!2,L\!-\!2)$. This results in a reduction of $\Delta\omega_D$ by a
factor $|(\omega_1-\omega_2)/(\omega_1+\omega_2)|$. Among a number of possible two-photon transitions,
a particularly strong signal is expected for
$(n,L)\!=\!(36,34)\!\to\!(34,32)$ (Fig.~\ref{energydig}), as this involves a large antiproton population  \cite{mhori2002,mhori2004}
in the resonance parent state $(36,34)$.
Whereas the states $(36,34)$ and $(35,33)$ are metastable with 1-$\mu{\rm s}$-scale lifetimes, the resonance daughter
state $(34,32)$ is Auger-dominated with the lifetime $\tau\sim 4$ ns \cite{Hori01,Kor03,yamaguchi2002,yamaguchi2004} that
corresponds to the natural width of a spectral line of $\sim\!30$ MHz.

The probability of inducing the transition can be enhanced \cite{hori2000,bjorkholm,salomaa1975,salomaa1976,salomaa1977,bordo,fort95,wei98}
by tuning $\omega_1$ and $\omega_2$ so that the virtual intermediate state of the two-photon
transition lies close (e.g. $|\Delta\omega_d/2\pi|<$ 10--20 GHz) to the real state $(35,33)$ such that,
\begin{equation}
\begin{array}{@{}l}
\displaystyle
\omega_1{(\rm SI)}=4\pi R_{\infty}c\left[E_{(35,33)}-E_{(34,32)}{(\rm a.u.)}\right]+\Delta\omega_d,
\\[4mm]\displaystyle
\omega_2{(\rm SI)}=4\pi R_{\infty}c\left[E_{(36,34)}-E_{(35,33)}{(\rm a.u.)}\right]-\Delta\omega_d.
\end{array}
\end{equation}
Here $R_{\infty}c=3.289842\!\times\!10^{15}$ Hz denotes the Rydberg constant,
and $E_{(n,L)}$ the binding energy of the $\overline{p}{\rm ^4He}^+$ state $(n,L)$ in atomic units.
At experimental conditions where this offset is much larger than the Doppler width ($|\Delta\omega_d|\gg\Delta\omega_D$),
the two-photon transition is expected to directly transfer the antiprotons populating the parent state $(36,34)$ to
the daughter state $(34,32)$, whereas the population in the intermediate state $(35,33)$ will be unaffected.

This paper is organized in the following way. Some details of the numerical methods are described in Section~\ref{detailcalc}.
The transition amplitude of the two-photon resonance $(36,34)$$\rightarrow$$(34,32)$ of $\overline{p}{\rm ^4He}$,
and the polarizabilities of the parent and daughter states $(36,34)$ and $(34,32)$ at various
offsets $\Delta\omega_d$ are estimated in Sections~\ref{tme} and \ref{cpol}. We next calculate the "background" polarizability due to
the contributions of $\overline{p}{\rm He}^+$ states other than the resonant intermediate state $(35,33)$ (Section~\ref{cnon}).
Based on these results, the ac Stark shift and broadening are semi-analytically estimated in Section~\ref{trf}. We next discuss the hyperfine structure
in two-photon transitions of the $\overline{p}{\rm ^4He}^+$ and $\overline{p}{\rm ^3He}^+$ isotopes (Section~\ref{hfss})
\cite{pask09,bakahfs,yam01,kinohfs,Kor06,bakalov07}. We numerically simulate the profile of several two-photon resonances
(Section~\ref{ore}) before concluding the paper. Atomic units (a.u.) are used to evaluate the state polarizabilities and transition amplitudes
\cite{khadjavi,chung1992,bonin,bhatia,michel,masili}, whereas International System of Units (SI) are used for
the transition frequencies, rates, and laser intensities relevant for future spectroscopy experiments.

\section{Details of the calculation}
\label{detailcalc}

For simplicity we take the linearly polarized laser field aligned along the
$z$-axis, such that the perturbation Hamiltonian in a laser field of frequency $\omega$ and amplitude $F$ has the form,
\begin{equation}
H' = -e\,\mathbf{F}(t)\,\mathbf{d},
\qquad
\mathbf{F}(t)=\mathbf{e}_zF\cos(wt).
\end{equation}
Here $\mathbf{d}=\sum_{a=1}^3 Z_a\mathbf{R}_a$ is the electric dipole
moment operator. The second order correction $E^{(2)}$ to the
unperturbed eigenenergy $E_0$ of a $\overline{p}{\rm He}^+$ state
vector $|0\rangle$ may then be expressed as,
\begin{equation}
E^{(2)} = -\,\frac{1}{2}\>\alpha_d^{zz}(\omega,M^2)\cdot F^2,
\label{secd}
\end{equation}
where $a_d^{ij}(\omega)$ is a tensor of the dynamic dipole polarizability:
\begin{equation}
\alpha_d^{ij} = -\sum_q
   \left[
      \frac{\langle 0|d^i|q \rangle\langle q|d^j|0 \rangle}{E_0-E_q+\omega}
      +\frac{\langle 0|d^i|q \rangle\langle q|d^j|0 \rangle}{E_0-E_q-\omega}
   \right].
\end{equation}
The energy of a $\overline{p}{\rm He}^+$ state vector $|q\rangle$ is denoted by $E_q$,
and the summation of $q$ is over all states which are accessible via single-photon
transition from the resonance parent state $|0\rangle$.

The tensor $a_d^{ij}(\omega)$ may be rewritten in terms of the irreducible scalar and tensor
polarizability operators,
\begin{equation}
\begin{array}{@{}l}
\displaystyle
\alpha_d^{ij} = \alpha_s+\alpha_t
   \left[
      \hat{L}^i\hat{L}^j+\hat{L}^j\hat{L}^i-\frac{2}{3}\mathbf{\hat{L}}^2
   \right],
\\[3mm]\displaystyle
\alpha_s^{} = \frac{1}{3}\bigl[a_++a_0+a_-\bigr],
\\[3mm] \displaystyle
\alpha_t^{} = -\frac{a_+L(2L\!-\!1)}{3(L\!+\!1)(2L\!+\!3)}
              +\frac{a_0(2L\!-\!1)}{3(L\!+\!1)}-\frac{a_-}{3}\>,
\end{array}
\label{ddp}
\end{equation}
where the angular momentum operator is denoted by $\hat{L}$.
The coefficients $a_+$, $a_0$, and $a_-$ are defined as follows,
\begin{equation}
\begin{array}{@{}l}
\displaystyle
a_+= -\frac{2}{2L+1}\sum_q
   \frac{(E_0\!-\!E_q)\bigl|\langle 0L\|\mathbf{d}\|q(L\!+\!1) \rangle\bigr|^2}
                                            {(E_0-E_q)^2-\omega^2},
\\[4mm]\displaystyle
a_0=-\frac{2}{2L+1}\sum_q
   \frac{(E_0\!-\!E_q)\bigl|\langle 0L\|\mathbf{d}\|qL \rangle\bigr|^2}
                                            {(E_0-E_q)^2-\omega^2},
\\[4mm]\displaystyle
a_-= -\frac{2}{2L+1}\sum_q
   \frac{(E_0\!-\!E_q)\bigl|\langle 0L\|\mathbf{d}\|q(L\!-\!1) \rangle\bigr|^2}
                                            {(E_0-E_q)^2-\omega^2}.
\end{array}
\label{apm}
\end{equation}
Here $a_+$ and $a_-$ represent the contributions from antiproton transitions
to states of normal parity which change the orbital angular momentum quantum number
$L$ of the antiproton by 1 or $-1$. The contribution $a_0$
involves transitions to $\overline{p}{\rm He}^+$ states of anomalous parity in which the
$L$-value is unchanged and the 1s-electron is excited to, e.g., the $2p$ state.

For our analysis it is convenient to define "background" polarizabilities using the above equations, where the
dominant contribution from the intermediate state of the two-photon transition is subtracted.
For example, the corresponding scalar and tensor background polarizabilities
of state $(n,L)=(36,34)$ can be calculated as,
\begin{equation}
\begin{array}{@{}l}
\displaystyle
\beta_s=\alpha_s+\frac{2\left(E_0\!-\!E_i\right)}{3(2L\!+\!1)}\cdot
   \frac{~\bigl|\langle 0L\|\mathbf{d}\|i(L\!-\!1) \rangle\bigr|^2}
                                            {\left(E_0-E_i\right)^2-\omega^2}, \\[3mm] \displaystyle
\beta_t=\alpha_t- \frac{2(E_0\!-\!E_i)}{3(2L\!+\!1)}\cdot
   \frac{~\bigl|\langle 0L\|\mathbf{d}\|i(L-1)\rangle\bigr|^2}
                                            {(E_0-E_i)^2-\omega^2}.
\end{array}
\label{ddpt}
\end{equation}
Here $E_0$ and $E_i$ denote the energies of state $(36,34)$ and the intermediate state $(35,32)$.

The transition matrix element $\kappa_{L,L-2,M}$ of the $\overline{p}{\rm He}^+$ two-photon
transition $(n,L,M)$$\rightarrow$$(n\!-\!2,L\!-\!2,M)$ induced by two linearly-polarized laser beams of total frequency,
$w_1+w_2\approx 4\pi R_\infty c(E_0-E_1)$, can be calculated using the Wigner $3j$-symbols as,
\begin{equation}\label{eq:ta}
\begin{array}{@{}l}
\displaystyle
\kappa_{L,L\!-2,M}=-\!\left(
\begin{array}{@{\,}c@{\;}c@{\;}c@{\,}}
L\!-\!1  & 1  & L  \\
 M  & 0\  & -M
\end{array}
\right)
\left(
\begin{array}{@{\,}c@{\;\,}c@{\;\,}c@{\,}}
L\!-\!1  & 1  & L\!-\!2  \\
 M  & 0  & -M\
\end{array}
\right)\!\!\!
\\[5mm]\displaystyle\hspace{5mm}
\times \sum_q
\biggl[
   \frac{\langle 1(L\!-\!2)\|\mathbf{d}\|q(L\!-\!1) \rangle
         \langle q(L\!-\!1)\|\mathbf{d}\|0L\rangle}{E_0-E_q-\omega_2}
\\[4mm] \displaystyle\hspace{14mm}
   +\frac{\langle 1(L\!-\!2)\|\mathbf{d}\|q(L\!-\!1) \rangle
         \langle q(L\!-\!1)\|\mathbf{d}\|0L\rangle}{E_0-E_q-\omega_1}
\biggr],
\end{array}
\end{equation}
wherein $|1\rangle$ denotes the state vector of the resonance daughter state of energy $E_1$.
This is related to the two-photon Rabi oscillation frequency (in atomic units) of this laser transition via the equation,
\begin{equation}
\Omega_{2\gamma M}{\rm (a.u.)}=\frac{1}{2}\left|\kappa_{L,L-2,M}\right|F_1F_2.
\label{oscr}
\end{equation}
The last term in Eq.~(\ref{eq:ta}) can be neglected at small offset frequencies $\Delta w_d$.

In order to calculate these quantities we must evaluate the reduced
matrix elements for the dipole operator $\mathbf{d}$ and diagonalize
the Hamiltonian. For this we use the variational exponential expansion
described in Ref.~\cite{Kor00}. The wave function for a state with a total
orbital angular momentum $L$ and of a total spatial parity $\pi=(-1)^L$ is
expanded as follows,
\begin{equation}\label{korowave}
\begin{array}{@{}l}
\displaystyle \Psi_{LM}^\pi(\mathbf{R},\mathbf{r}_1) =
       \sum_{l_1+l_2=L}
         \mathcal{Y}^{l_1l_2}_{LM}(\hat{\mathbf{R}},\hat{\mathbf{r}}_1)
         G^{L\pi}_{l_1l_2}(R,r_1,r_2),
\\[4mm]\displaystyle
G_{l_1l_2}^{L\pi}(R,r_1,r_2) = \sum_{n=1}^N \Big\{C_n\,\mbox{Re} \bigl[e^{-\alpha_n R-\beta_n r_1-\gamma_n r_2}\bigr]
\\[2mm]\displaystyle\hspace{30mm}
+D_n\,\mbox{Im} \bigl[e^{-\alpha_n R-\beta_n r_1-\gamma_n r_2}\bigr] \Big\}.
\end{array}
\end{equation}
where the complex exponents $\alpha$, $\beta$, and $\gamma$ are generated in
a pseudorandom way, $\mathbf{R}$ and $\mathbf{r}_1$ are position vectors
of an antiproton and an electron with respect to a helium nucleus, and
$r_2$ the distance between the antiproton and electron. Further details may be found in Refs.~\cite{Kor00,Kor03}.

This method has been previously employed to calculate the nonrelativistic energies
of $\overline{p}{\rm He}^+$ with a relative precision of around 1 part in $10^{12}$
\cite{Kor00,Kor03,Kor08}. We here determined the nonrelativistic values of the
transition matrix elements using the same wavefunctions to a precision of
$\sim 10^{-6}$. This was more than adequate for our purpose of roughly estimating the two-photon
transition probabilities and ac Stark shifts relevant to future spectroscopy experiments.
The $\overline{p}{\rm He}^+$ states of unnatural (or anomalous) parity $(-1)^{L+1}$  involve an electron
in an excited state and therefore are not metastable \cite{korobov96}. They will not be considered here.

\section{Results}

\subsection{Transition matrix element}
\label{tme}

We first calculated the transition amplitude $\kappa_{L,L-2,M}$ for the two-photon resonance $(n,L)=(36,34)$$\rightarrow$$(34,32)$ in $\overline{p}{\rm ^4He}^+$
at various offsets $\omega_d$ from the intermediate $(35,33)$ state, and estimated the laser intensities needed to drive this transition.
In Fig.~\ref{energydig2}, sequences of single-photon transitions connecting the states
$(36,34)$ and $(34,32)$ are indicated by straight arrows, together with the corresponding dipole moment
$|\langle 0L\|\mathbf{d}\|q(L-1)\rangle|$. The atomic units shown here can be converted to SI units, the corresponding spontaneous decay rates in s$^{-1}$ obtained using the equation,
\begin{equation}
\gamma {\rm (SI)}=\frac{e^2a_0^2}{4\pi\epsilon_0}\frac{4\omega^3_{q0}}{3\hbar c^3}\frac{|\langle 0L\|\mathbf{d}\|q(L-1)\rangle|^2 {\rm (a.u.)}}{2L+1}.
\label{decsi}
\end{equation}
The SI-unit constants that appear in the above equation are, $e$: the elementary charge, $a_0$: the Bohr radius,
$\epsilon_0$: the dielectric constant of vacuum, $\hbar$: the reduced Planck constant, and $c$ the speed of light.
The angular transition frequency between states $|q\rangle$ and $|0\rangle$ are denoted by $\omega_{q0}$.

\begin{figure}[htbp]
\begin{center}
\vspace{-3mm}
\includegraphics[width=80mm]{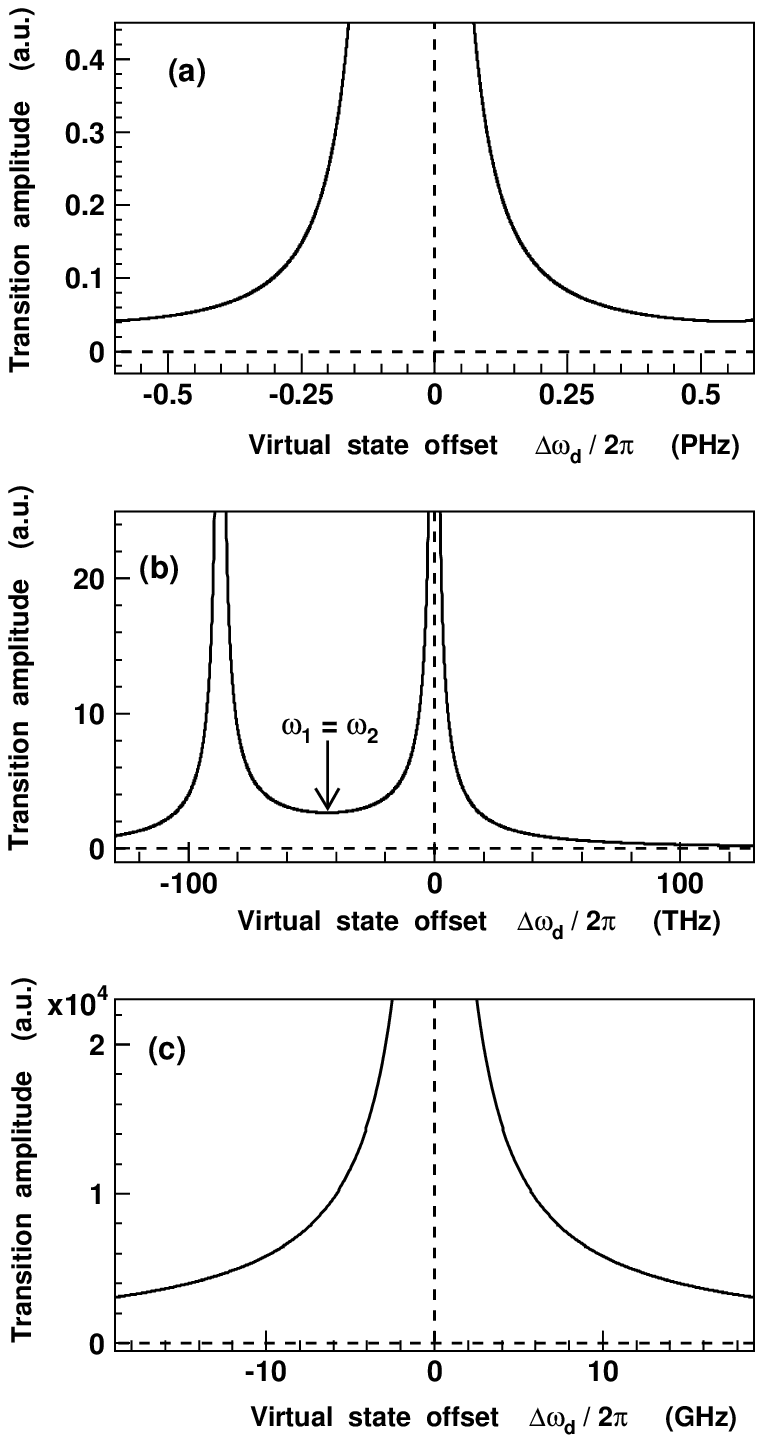}
\vspace{-3mm}
\end{center}
\caption{\label{twotrans} Transition matrix amplitude $|\kappa_{L,L-2}|$ of the two-photon resonance
$(n,L)=(36,34)$$\rightarrow$$(34,32)$ as a function of the offset
$\Delta\omega_d/2\pi$ in the two laser frequencies $\omega_1$ and $\omega_2$ from the virtual intermediate state.
Transition probability is averaged over the magnetic quantum number $M$, see text.
Arrow indicates the position where the two laser beams have equal frequencies.}
\end{figure}

Two types of transitions $(n,L)\rightarrow(n,L\!-\!1)$ and $(n,L)$$\rightarrow$$(n\!-\!1,L\!-\!1)$ have the largest amplitudes
of $\sim 1$ a.u., but the latter kind which conserve the vibrational quantum number $v=n\!-\!L\!-\!1$ and involve fluorescence
photons of frequency $\omega_{q0}/2\pi\sim 10^{15}$ Hz constitute the dominant channels of spontaneous decay. These transitions are most favorable for laser spectroscopy.
The transition frequencies, dipole moments, and decay rates of some single-photon resonances in $\overline{p}{\rm ^4He}^+$ and $\overline{p}{\rm ^3He}^+$ of the type $(n,L)$$\rightarrow$$(n\!-\!1,L\!-\!1)$ are shown
in Table~\ref{dipo}. The dipole moments for the higher-lying infrared transitions involving states with $n\sim 40$ are
relatively large ($>2$ a.u.), whereas for UV transitions in the $n\le 33$ regions it is reduced to $<1$ a.u.
On the other hand, the radiative decay rates increase for lower-$n$ transitions, e.g. from $4.7\times 10^5$ s$^{-1}$
for $(40,36)$$\rightarrow$$(39,35)$, to $6.6\times 10^{5}$ s$^{-1}$ for $(32,31)$$\rightarrow$$(31,30)$
due to the $\omega_{q0}^3$-dependence.

\begin{table*}[htbp]
\caption{\label{dipo}
Some single-photon transitions in $\overline{p}{\rm ^4He}^+$ and $\overline{p}{\rm ^3He}^+$:
their vibrational quantum number $v$, transition frequencies, dipole moments, and radiative rates.}
\begin{ruledtabular}
\begin{tabular}{ccccc}
$(n,L)\rightarrow (n^{\prime},L^{\prime})$ & $v$ & Trans. freq. & Dipole moment & Rad. rate \\
& & (THz) & (a.u.) & ($10^5$ ${\rm s}^{-1}$) \\ \colrule
\multicolumn{5}{c}{$\overline{p}{\rm ^4He}^+$ states} \\
$(40,36)\rightarrow(39,35)$ & 3 & 444.8 & 2.28 &  4.71 \\
$(39,35)\rightarrow(38,34)$ & 3 & 501.9 & 2.02 & 5.44 \\
$(38,35)\rightarrow(37,34)$ & 2 & 566.1 & 1.82 & 6.36 \\
$(37,34)\rightarrow(36,33)$ & 2 & 636.9 & 1.58 & 7.01 \\
$(37,35)\rightarrow(36,34)$ & 1 & 638.6 & 1.61 & 7.15 \\
$(36,34)\rightarrow(35,33)$ & 1 & 717.5 & 1.38 & 7.63\\
$(35,33)\rightarrow(34,32)$ & 1 & 804.6 & 1.16 & 7.92\\
$(33,32)\rightarrow(32,31)$ & 0 & 1012.4 & 0.79 & 7.42 \\
$(32,31)\rightarrow(31,30)$ & 0 & 1132.6 & 0.62 & 6.62 \\
\multicolumn{5}{c}{$\overline{p}{\rm ^3He}^+$ states} \\
$(39,35)\rightarrow(38,34)$ & 3 & 445.8 & 2.30    & 4.96\\
$(38,34)\rightarrow(37,33)$ & 3 & 505.2 & 2.03    & 5.77\\
$(37,34)\rightarrow(36,33)$ & 2 & 572.0 & 1.83    & 6.81\\
$(36,33)\rightarrow(35,32)$ & 2 & 646.2 & 1.58    & 7.55\\
$(35,33)\rightarrow(34,32)$ & 1 & 730.8 & 1.37    & 8.27\\
$(34,32)\rightarrow(33,31)$ & 1 & 822.8 & 1.15    & 8.59\\
$(33,32)\rightarrow(32,31)$ & 0 &  928.8  &  0.95 & 8.44\\
$(32,31)\rightarrow(31,30)$ & 0 & 1043.1 &  0.77 & 7.94\\
\end{tabular}
\end{ruledtabular}
\end{table*}

Using the single-photon dipole moments calculated above, we derive the two-photon
transition amplitude $\kappa_{L,L-2,M}$ of the resonance $(36,34)$$\rightarrow$$(34,32)$
in $\overline{p}{\rm ^4He}^+$ for cases where the virtual intermediate state is offset
over a large range between $\Delta\omega_d/2\pi=-0.6$ and 0.6 PHz from the state $(35,33)$.
The atom is excited by two linearly-polarized, counterpropagating laser beams.
Fig.~\ref{twotrans} (a)--(c) show the amplitude $|\kappa_{L,L-2}|$ averaged over all $\sim 70$ transitions
between the magnetic substates which conserve the $M$-value,
\begin{equation}
\left|\kappa_{L,L-2}\right|^2=\frac{1}{2L+1}\sum_M\left|\kappa_{L,L-2,M}\right|^2.
\end{equation}
The $|\kappa_{L,L-2}|$-values are usually small, e.g. a few a.u. for lasers of equal frequency
($\omega_1=\omega_2$). This is smaller than the amplitude $\sim 7.8$ a.u. \cite{Haas2006}
for the 1s-2s two-photon transition of atomic hydrogen excited by 243-nm laser light.
Gigawatt-scale laser intensities would be needed to induce the antiprotonic transition within the microsecond-scale
lifetime of $\overline{p}{\rm He}^+$.  On the other hand, the transition probabilities can be strongly enhanced to $\gg10^3$ a.u.
by tuning the virtual intermediate state within $\sim 20$ GHz of the real states $(n,L)=(34,33)$, $(35,33)$, or $(36,33)$.
Eq.~\ref{oscr} indicates that the transition can then be induced using nanosecond laser pulses
of electric field $F\sim (1-2)\times 10^{-6}$ a.u. According to the equation,
\begin{equation}
I ({\rm SI})=\frac{1}{2}\epsilon_0c\left(\frac{eF(\rm a.u.)}{4\pi\epsilon_0a_0^2}\right)^2,
\end{equation}
this corresponds to a peak intensity of $I\sim 10^4$--$10^5$ W/cm$^2$ which is achievable
using titanium sapphire lasers of narrow linewidth \cite{horiopt09}. The second maximum at
$\omega_d/2\pi\sim -85$ THz in Fig.~\ref{twotrans} (b) corresponds to the case of the
$\omega_1$ and $\omega_2$ lasers resonating with the respective transitions
$(36,34)$$\rightarrow$$(35,33)$ and $(35,33)$$\rightarrow$$(34,32)$.

\subsection{Polarizabilities}
\label{cpol}

\begin{figure*}[htbp]
\begin{center}
\vspace{-3mm}
\includegraphics[width=160mm]{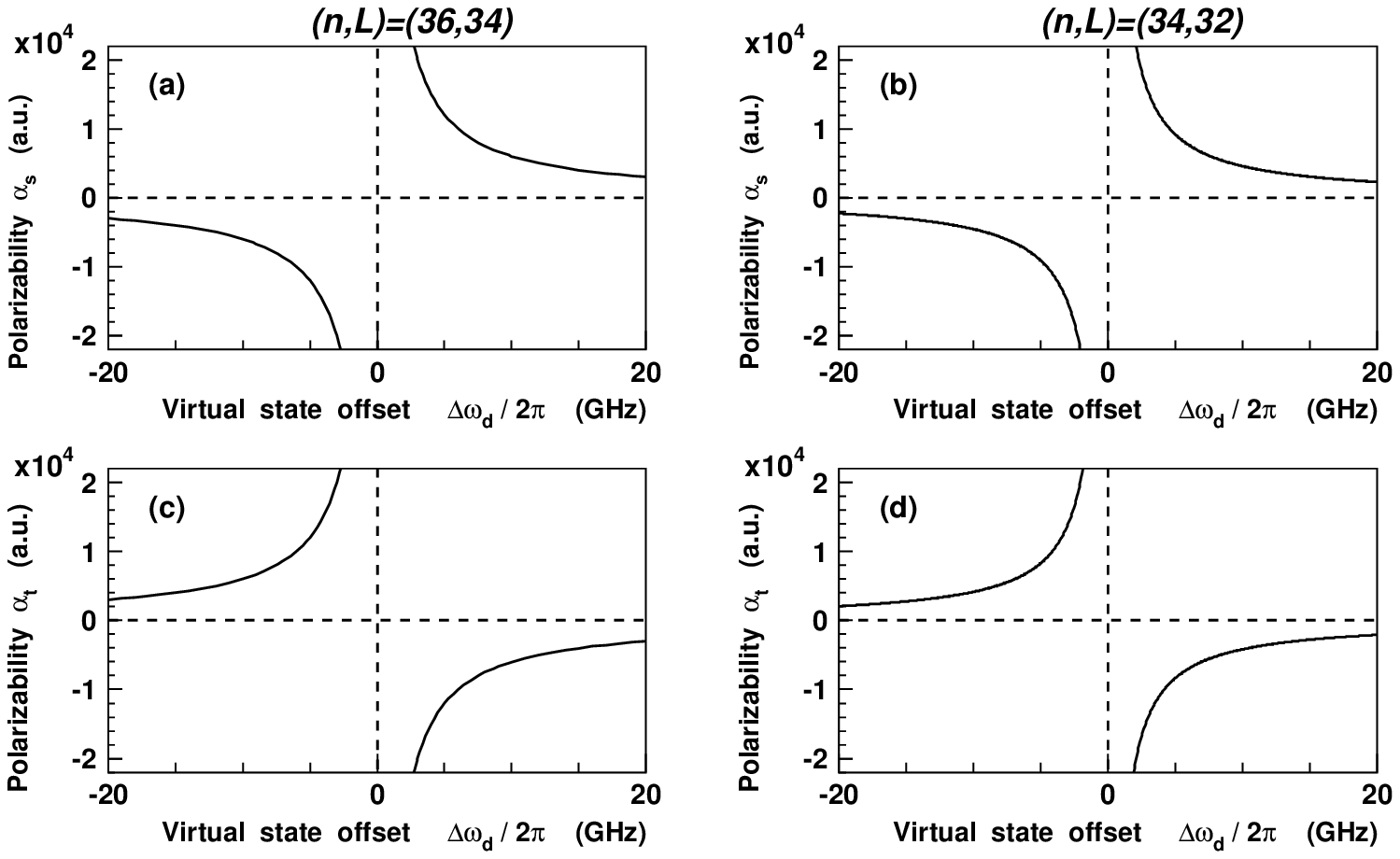}
\vspace{-3mm}
\end{center}
\caption{\label{dynpol_s} Scalar dipole polarizability of states $(n,L)=(36,34)$ and $(34,32)$
versus frequency offset $\Delta\omega_d/2\pi$ of the virtual intermediate state from the
real state $(35,33)$:
$\alpha_s(w_1)_{(34,32)}$ and $\alpha_s(w_2)_{(36,34)}$
as defined in Eqs.~(\ref{ddp}) (a)--(b). Tensor polarizabilities
$\alpha_t(w_1)_{(34,32)}$ and $\alpha_t(w_2)_{(36,34)}$ (c)--(d).}
\end{figure*}

We next evaluate the polarizabilities of the parent and daughter states of the transition.
In Figs.~\ref{dynpol_s} (a) and (c), the scalar and tensor polarizability components, $\alpha_s(\omega_1)_{(34,32)}$,
$\alpha_t(\omega_1)_{(34,32)}$, of state $(n,L)=(34,32)$ are shown. To simplify the calculation, we initially assume that
the atom is irradiated by a single laser field of frequency $\omega_1$ (see Fig.~\ref{energydig}) corresponding to offsets
between $\Delta\omega_d/2\pi=-20$ and 20 GHz from the state $(35,33)$.
A similar plot for the polarizabilities $\alpha_s(\omega_2)_{(36,34)}$
and $\alpha_t(\omega_2)_{(36,34)}$ of state $(36,34)$ under irradiation by a laser field of $\omega_2$
are shown in Figs.~\ref{dynpol_s} (b) and (d).

As the laser frequencies are offset from $\Delta\omega_d/2\pi=-100$ to -6 GHz,
the scalar polarizabilities decrease from $\sim -500$ to $\sim -1\times 10^4$ a.u.,
whereas the tensor polarizabilities have opposite sign and increase from $\sim 500$ to
$\sim 1\times 10^4$ a.u. (Table ~\ref{shift}). These polarizabilities of 1000--10000 a.u. correspond to
50--500 Hz/(W/cm$^2$) in SI units according to the equation,
\begin{equation}
\alpha ({\rm a.u.})=\alpha ({\rm SI})\times\frac{\hbar^2}{m_ea_0^4\alpha}.
\end{equation}
The graphs follow a reciprocal $\Delta\omega_d^{-1}$ dependence and are approximately symmetric with
respect to the origin. This simple behavior suggests that the ac Stark shift is primarily caused by the contribution
from the intermediate state $(35,33)$.

\subsection{Contribution of nonresonant states}
\label{cnon}

\begin{table*}[htbp]
\caption{\label{shift}
Scalar and tensor polarizabilities of the $\overline{p}{\rm ^4He}^+$ states $(36,34)$ and $(34,32)$ at
offset frequencies of the lasers between $\Delta\omega_d/2\pi=-100$ and 100 GHz, see text.}
\begin{ruledtabular}
\begin{tabular}{ccccccc}
& \multicolumn{6}{c}{Polarizabilities (a.u.)} \\
 $\Delta\omega_d/2\pi$  &   $-100$ GHz  & $-12$ GHz  & $-6$ GHz & 6 GHz & 12 GHz & 100 GHz \\  \colrule
    \multicolumn{7}{c}{$(n,L)=(36,34)$ state} \\
 $\alpha_s(\omega_2)_{(36,34)}$ & $-602$ & $-5.02\times 10^3$ & $-1.00\times 10^4$ & $1.00\times 10^4$&  $5.02\times 10^{3}$ & $602$ \\
 $\alpha_s(\omega_1)_{(36,34)}$ & $-1.61$ & $-1.61$ & $-1.61$ & $-1.61$ &  $-1.61$ & $-1.61$ \\
 $\alpha_t(\omega_2)_{(36,34)}$ & $601$ &$5.02\times 10^{3}$ & $1.00\times 10^4$ & $-1.00\times 10^4$ & $-5.02\times 10^{3}$ & $-603$ \\
 $\alpha_t(\omega_1)_{(36,34)}$ & $0.204$ & $0.203$ & $0.203$ & $0.203$ & $0.203$ & $0.203$ \\
 $\beta_s(\omega_2)_{(36,34)}$ & $-0.403$ & $-0.401$ &  $-0.401$ & $-0.401$ &  $-0.401$ & $-0.400$\\
 $\beta_t(\omega_2)_{(36,34)}$ & $-0.945$ & $-0.946$ & $-0.946$ & $-0.947$ & $-0.947$ & $-0.948$\\
   \multicolumn{7}{c}{$(n,L)=(34,32)$ state} \\
 $\alpha_s(\omega_2)_{(34,32)}$ & $-2.06$ &  $-2.06$& $-2.06$ &  $-2.06$& $-2.06$ &  $-2.06$\\
 $\alpha_s(\omega_1)_{(34,32)}$ & $-459$ & $-3.81\times 10^{3}$ & $-7.62\times 10^3$ &  $7.62\times 10^3$& $3.81\times 10^{3}$ & $456$\\
 $\alpha_t(\omega_2)_{(34,32)}$ & $0.265$ & $0.265$& $0.265$ & $0.265$ & $0.265$ & $0.264$\\
 $\alpha_t(\omega_1)_{(34,32)}$ & $416$ & $3.47\times 10^{3}$ & $6.94\times 10^3$ & $-6.95\times 10^3$ & $-3.48\times 10^{3}$ & $-417$\\
 $\beta_s(\omega_1)_{(34,32)}$ & $-1.36$ & $-1.36$ &  $-1.36$ & $-1.36$ &  $-1.36$ & $-1.36$\\
 $\beta_t(\omega_1)_{(34,32)}$ & $-0.401$ & $-0.401$ & $-0.401$ & $-0.401$ & $-0.402$ & $-0.402$\\
\end{tabular}
\end{ruledtabular}
\end{table*}

\begin{figure*}[htbp]
\begin{center}
\vspace{-3mm}
\includegraphics[width=160mm]{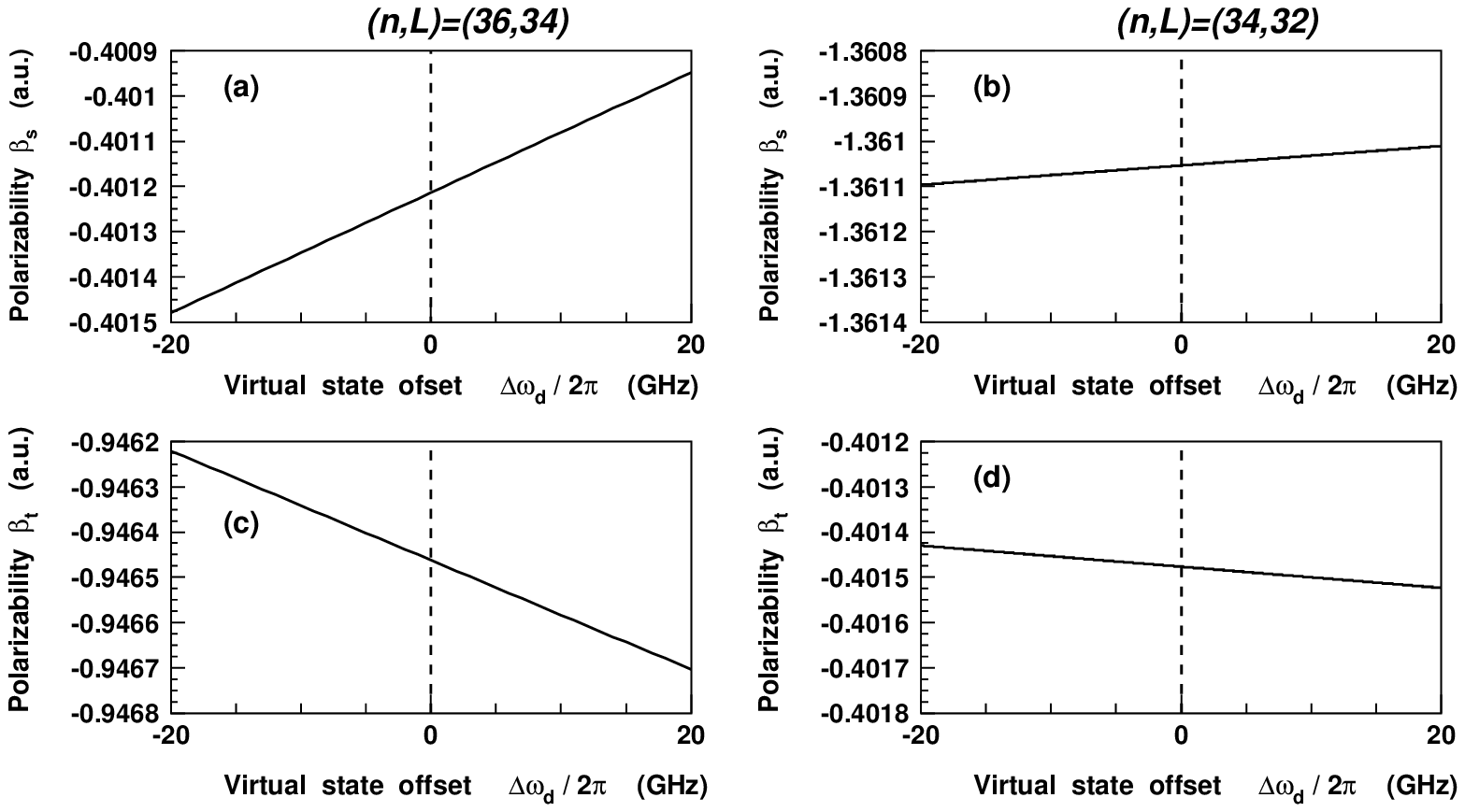}
\vspace{-3mm}
\end{center}
\caption{\label{dynpol_t} Residual scalar dipole polarizability of states
$(n,L)=(36,34)$ and $(34,32)$ versus frequency offset $\Delta\omega_d/2\pi$
of the virtual intermediate state from the real state $(35,33)$:
$\beta_s(\omega_1)_{(34,32)}$ and $\beta_s(\omega_1)_{(34,32)}$, and
$\beta_t(\omega_2)_{(36,34)}$ and $\beta_t(\omega_2)_{(36,34)}$.
The contributions of all states except the intermediate one $(35,33)$ are included.}
\end{figure*}

We next study the non-resonant or "background" contributions (Eq.~\ref{ddpt}) to the polarizabilities
from all $\overline{p}{\rm He}^+$ states other than the intermediate state $(n,L)=(35,33)$.
This will allow us to estimate how far the measured ac-Stark shift would deviate from
the predictions of a simple three-level model which include only the resonance parent, daughter,
and intermediate states. Figs.~\ref{dynpol_t} (a) and (c) show the background scalar and tensor polarizabilities
$\beta_s(\omega_2)_{(36,34)}$ and $\beta_t(\omega_2)_{(36,34)}$
of state $(36,34)$ when irradiated with a single laser field
of frequency $\omega_2$ at offsets between $\Delta\omega_d/2\pi=-20$ and 20 GHz.
They remained relatively constant at $\sim\!-0.40$ and $0.95$ a.u. respectively (Table~\ref{shift}).
Figs.~\ref{dynpol_t} (b) and (d) are the corresponding plots of $\beta_s(\omega_1)_{(34,32)}$
and $\beta_t(\omega_1)_{(34,32)}$ for state $(34,32)$ irradiated with the $\omega_1$-laser.
They are similarly constant ($-1.36$ and $-0.40$ a.u.). All these background polarizabilities are
at least three orders of magnitude smaller than the dominant contributions to $\alpha_s$ and $\alpha_t$
arising from the intermediate state $(35,33)$ at small offsets $|\Delta\omega_d/2\pi|<12$ GHz.

The case of two counterpropagating laser fields of angular frequencies $\omega_1$
and $\omega_2$, and amplitudes $F_1$ and $F_2$ irradiating the atom simultaneously
will next be considered. The perturbation Hamiltonian for this can be expressed as,
\begin{equation}
\begin{array}{@{}l}
H' = -e\,\mathbf{F}(t)\,\mathbf{d},
\\[2mm]
\mathbf{F}(t)=\mathbf{e}_z\left[F_1\cos(w_1t)\!+\!F_2\cos(w_2t)\right].
\end{array}
\end{equation}
As shown in Ref.~\cite{Hansch86},
the interference effect between the two laser fields can be neglected in the case
of $\omega_1\neq\omega_2$. The contribution to the ac Stark shift of state $(36,34)$
from the $\omega_1$-laser [which is far off-resonance with respect to the upper
single-photon transition $(36,34)$$\rightarrow$$(35,33)$] is expressed by the scalar
and tensor polarizabilities $\alpha_s(\omega_1)_{(36,34)}$ and
$\alpha_t(\omega_1)_{(36,34)}$. The calculated values
at offsets between $\Delta\omega_d/2\pi= -20$ and 20 GHz were respectively
$-1.6$ and $-0.2$ a.u. (Table~\ref{shift}). The corresponding values for the daughter state
$\alpha_s(\omega_2)_{(34,32)}$ and $\alpha_t(\omega_2)_{(34,32)}$ were
also small ($-2.1$ and $0.3$ a.u.).

We conclude that the two-photon spectroscopy experiment depicted in Fig.~\ref{energydig}
can be accurately simulated by a simple model involving three states interacting with two laser beams.
Any non-resonant contribution from other $\overline{p}{\rm He}^+$ states are at least
three orders of magnitude smaller.

\subsection{ac Stark shifts of transition frequency}
\label{trf}

\begin{figure*}[htbp]
\begin{center}
\vspace{-3mm}
\includegraphics[width=160mm]{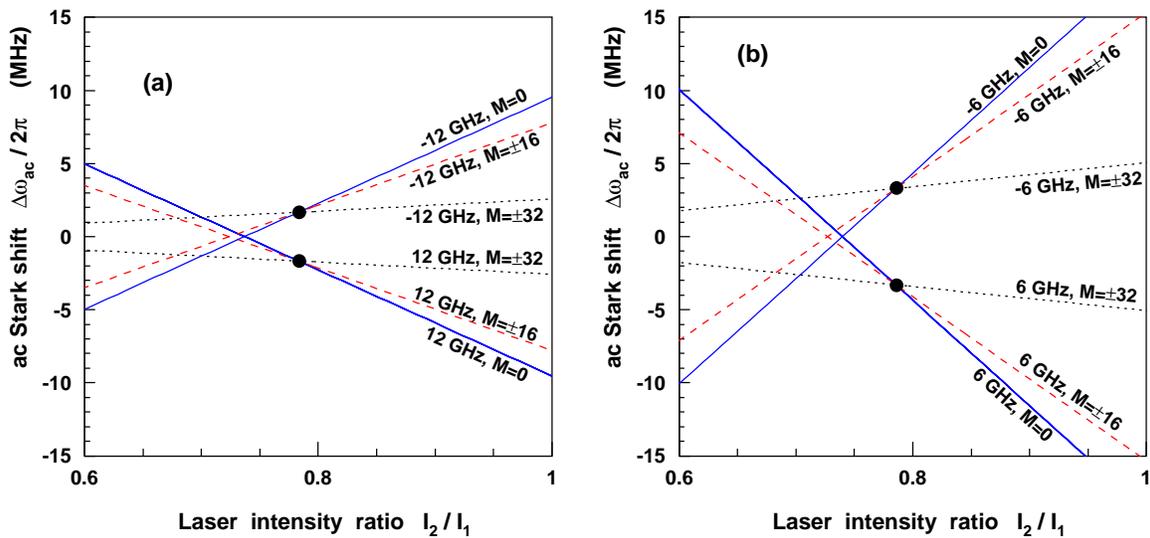}
\vspace{-3mm}
\end{center}
\caption{\label{starkcomp} (Color online) The ac Stark shifts in the $\overline{p}{\rm ^4He}^+$ resonance
$(36,34)$$\rightarrow$$(34,32)$ estimated from the state polarizabilities $\alpha_s$ and $\alpha_t$,
as a function of the intensity ratio $I_2/I_1$ between the two laser beams with $I_1$ fixed at $5\times 10^4$ W/cm$^2$.
Shifts for transitions involving the magnetic substates $M=0$, $\pm 16$, and $\pm 32$
are indicated. Virtual intermediate state is tuned $\Delta\omega_d/2\pi=\pm 12$ GHz (a) and
$\pm 6$ GHz (b) from the state $(35,33)$.}
\end{figure*}

The ac Stark shift in the transition frequency of the $(36,34)$$\rightarrow$$(34,32)$ resonance
can be analytically estimated as,
\begin{equation}
\frac{\Delta\omega_{\rm ac,M}}{2\pi}= 2R_{\infty}c\left[E^{(2)}_{(36,34,M)}-E^{(2)}_{(34,32,M)}\right],
\label{shiftac}
\end{equation}
wherein the ac Stark shifts $E^{(2)}_{(36,34,M)}$ and $E^{(2)}_{(34,32,M)}$ in the parent and daughter states of magnetic
substate $M$ induced by the two linearly-polarized laser fields can be approximated as,
\begin{equation}
\begin{split}
\displaystyle
E^{(2)}_{(36,34,M)}&\sim -\frac{1}{2}\bigg[\alpha_s(\omega_2)_{(36,34)} \\
&+\frac{3M^2-1190}{2278}\alpha_t(\omega_2)_{(36,34)}\bigg]F_2^2,
\\[3mm]\displaystyle
E^{(2)}_{(34,32,M)} & \sim -\frac{1}{2}\bigg[\alpha_s(\omega_1)_{(34,32)} \\
&+\frac{3M^2-1056}{2016}\alpha_t(\omega_1)_{(34,32)}\bigg]F_1^2.
\end{split}
\end{equation}

In Fig.~\ref{starkcomp} (a), the ac Stark shift $\Delta\omega_{\rm ac,M}/2\pi$ for
$M$-values $0$, $\pm 16$, and $\pm 32$, and two laser offsets $\Delta\omega_d /2\pi=-12$ and 12 GHz
obtained from the above equations are plotted as a function of the intensity ratio
$I_2/I_1=F_2^2/F_1^2$ between the two laser beams. Here $I_1$ is fixed at
$\sim 5\times 10^4$ W/cm$^2$ while $I_2$ is scanned between
$(3-5)\times 10^4$ W/cm$^2$.
We find a positive shift ($\Delta\omega_{\rm ac}>0$) at two combinations of laser offsets
and intensities ($\Delta\omega_d>0$, $I_1\gg I_2$) and ($\Delta\omega_d<0$, $I_1\ll I_2$).
Conversely the shift is negative $\Delta\omega_{\rm ac}<0$ at
($\Delta\omega_d<0$, $I_1\gg I_2$) and ($\Delta\omega_d>0$, $I_1\ll I_2$).
In addition to the ac Stark shift, the tensor polarizabilities cause
the resonance line to split depending on the $M$-value of the involved states.
At conditions of $I_2/I_1<0.6$ or $>1$ the ac Stark shift and splitting can reach values of
more than 5--10 MHz. At smaller offsets $\left|\Delta\omega_d/ 2\pi\right|=6$ GHz, the
ac Stark shift and splitting become twice as large [Fig.~\ref{starkcomp} (b)].

The ac Stark shift arising from $\alpha_s$ and the splitting due to $\alpha_t$ can
be minimized by adjusting the laser intensities to the values indicated by filled circles in
Figs.~\ref{starkcomp} (a)--(b),
\begin{equation}
\frac{I_2}{I_1}\sim \frac{\alpha_s(\omega_1)_{(34,32)}}{\alpha_s(\omega_2)_{(36,34)}}\sim 0.76.
\label{sda}
\end{equation}
An important point is that when the sign of the laser detuning is reversed, e.g., from $\Delta\omega_d/2\pi=-12$ to 12 GHz,
the resulting ac Stark shift also reverses sign but its magnitude is the same,
\begin{equation}
\Delta\omega_{\rm ac}(\omega_1,\omega_2)=-\Delta\omega_{\rm ac}(\omega_1-2\Delta\omega_d,\omega_2+2\Delta\omega_d).
\end{equation}
This means that the residual ac Stark shift can be canceled by comparing the two-photon transition frequencies
measured at laser offsets of opposite sign but the same absolute value, i.e., $\Delta\omega_d$ and $-\Delta\omega_d$.

\begin{figure*}[htbp]
\begin{center}
\vspace{-3mm}
\includegraphics[width=160mm]{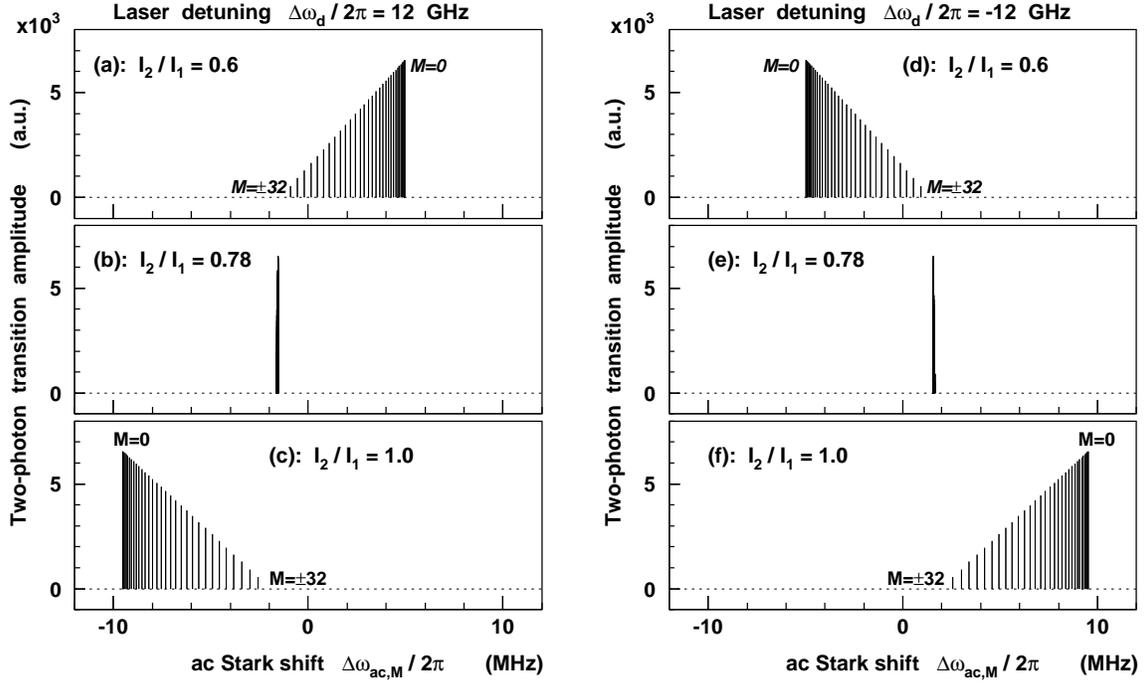}
\vspace{-3mm}
\end{center}
\caption{\label{linesm} Expected positions and transition amplitudes $\left|\kappa_{L,L-2,M}\right|$ of the M-sublines of the
two-photon transition $(36,34)$$\rightarrow$$(34,32)$ in $\overline{p}{\rm ^4He}^+$,
wherein the virtual intermediate state is offset $\Delta\omega_d/2\pi\sim 12$ GHz (a)--(c) and
$-12$ GHz (d)--(f) from the state $(35,33)$. The intensity ratios $I_2/I_1$ between the two
laser beams were varied between 0.6, 0.78, and 1.0 in each figure, whereas $I_1$ was kept
constant at $\sim 5\times10^{4}$ W/cm$^2$.}
\end{figure*}

Fig.~\ref{linesm} shows the ac Stark shifts and transition amplitudes $\left|\kappa_{L,L-2,M}\right|$ of
all magnetic sublines between $M=-32$ and $32$ of this two-photon resonance at laser offsets
$\Delta\omega_d/2\pi=12$ GHz (a)--(c) and $-12$ GHz (d)--(f). The shifts were calculated
at three ratios of the laser intensities $I_2/I_1=0.6$, 0.78, and 1.0 with $I_1$ fixed at
$5\times10^{4}$ W/cm$^2$. The ac Stark effect causes a triangular profile with the $M=0$
transitions being strongest and shifting the most. In actual experiments involving pulsed laser
beams, these shifts and splittings would smear out due to nanosecond-scale changes in the
light field intensities $I_1(t)$ and $I_2(t)$; the observed intensities of the magnetic sublines
would have a nonlinear dependence on $\left|\kappa_{L,L-2,M}\right|^2$, $I_1(t)$,
and $I_2(t)$ as simulated in Sec.~\ref{ore}.

\begin{figure*}[htbp]
\begin{center}
\vspace{-3mm}
\includegraphics[width=160mm]{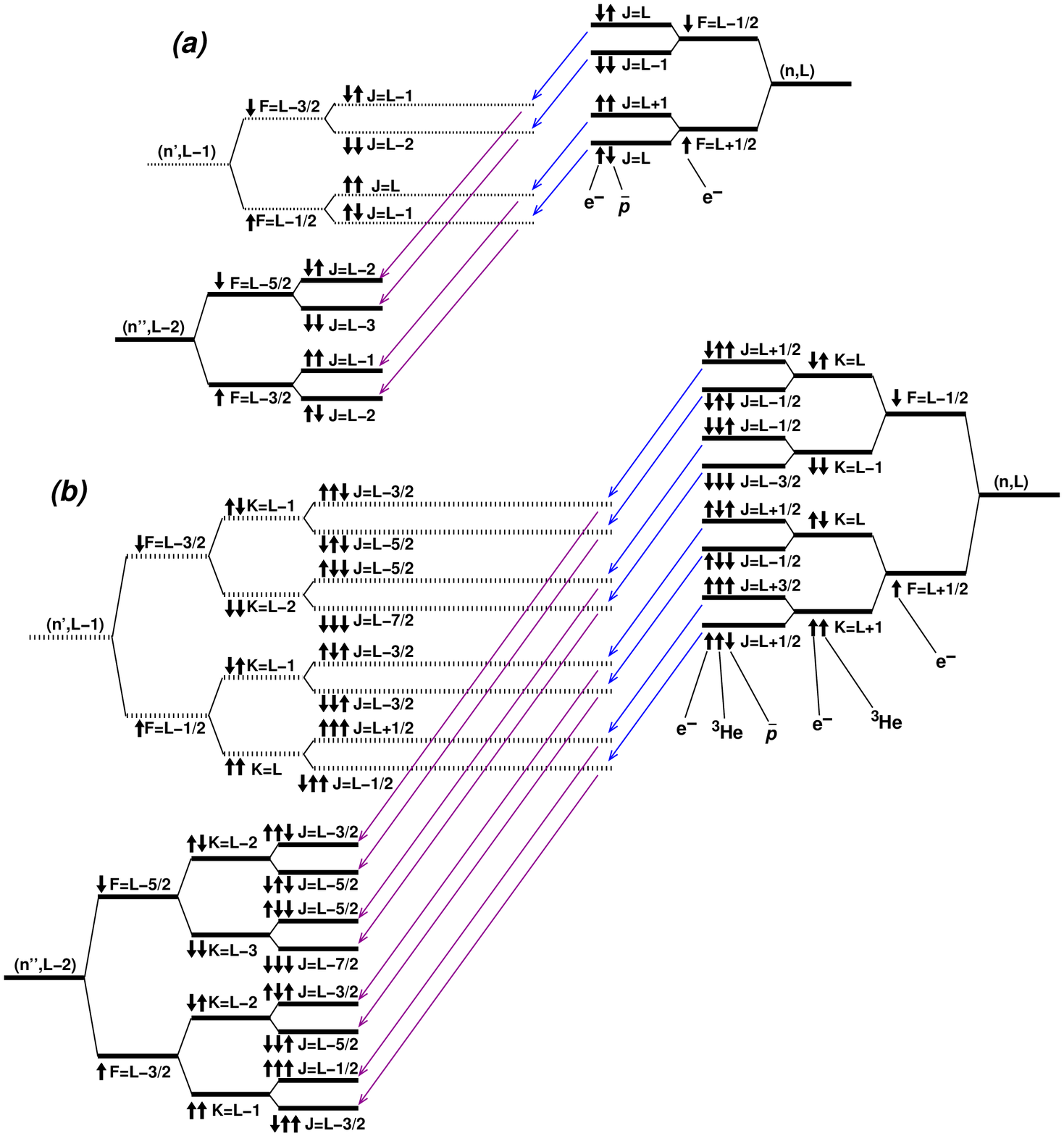}
\vspace{-3mm}
\end{center}
\caption{\label{hfstwotranshe} (Color online) Energy level diagram showing the hyperfine sublevels of the
resonance parent, intermediate, and daughter states involved in the two-photon transitions
of \protect{$\overline{p}{\rm ^4He}^+$} (a). The spin orientation \protect{$(S_{e},S_{\overline{p}})$} of each
hyperfine sublevel is indicated by arrows. The four strongest two-photon transitions are
indicated with solid arrows. The same figure in the case of $\overline{p}{\rm ^3He}^+$,
showing the eight strong lines between hyperfine levels $(S_{e},S_{h},S_{\overline{p}})$, see text.}
\end{figure*}

\subsection{Hyperfine structure}
\label{hfss}

We next study the hyperfine lines that appear in the two-photon resonance profile.
The hyperfine substates of the parent, intermediate, and daughter
states $(n,L)$, $(n-1,L-1)$, and $(n-2,L-2)$ in $\overline{p}{\rm ^4He}^+$ are shown schematically in
Fig.~\ref{hfstwotranshe} (a). Due to the dominant interaction between the electron spin $\mathbf{S}_{e}$
and the antiproton orbital angular momentum $\mathbf{L}$, a pair of fine structure sublevels of intermediate
angular momentum quantum number $F=L\pm 1/2$ and a splitting 10--15 GHz arise.
The interactions involving the antiproton spin $\mathbf{S}_{\overline{p}}$ cause
each fine structure sublevel to split by a few hundred MHz into pairs of hyperfine sublevels of total angular momentum quantum number $J=F\pm 1/2$. In Fig.~\ref{hfstwotranshe} (a), the spin orientations of the electron and antiproton $(S_{e},S_{\overline{p}})$ are indicated for the four hyperfine sublevels. For example the energetically highest-lying component consists of a spin-down electron and spin-up antiproton, i.e.\ $(S_{e},S_{\overline{p}})=(\downarrow\uparrow)$.

In the $\overline{p}{\rm ^3He}^+$ case [Fig.~\ref{hfstwotranshe} (b)], the
electronic fine structure sublevels of $F=L\pm1/2$ are similarly split by $\sim\! 10$ GHz.
Each fine structure sublevel is then split into pairs of $^3$He hyperfine sublevels of intermediate
angular momentum $K=F\pm\frac{1}{2}$ arising from the interactions
involving the nuclear spin $\mathbf{S}_h$. The antiproton spin gives rise to eight hyperfine sublevels of total
angular momentum $J=K\pm\frac{1}{2}$. The spin orientations of the
three constituent particles $(S_{e},S_h,S_{\overline{p}})$ are indicated
for each substate in Fig.~\ref{hfstwotranshe} (b).

In Fig.~\ref{hfstwotranshe} (a)--(b), the four and eight strongest two-photon transitions between the
hyperfine sublevels of $\overline{p}{\rm ^4He}^+$ and $\overline{p}{\rm ^3He}^+$
are indicated by arrows. These transitions pass through the virtual intermediate state without flipping
the spin of any constituent particle. Many other transitions are possible, but they all involve spin-flip
and so their amplitudes are suppressed by three orders of magnitude or more.

\subsection{Optical rate equations}
\label{ore}

To simulate $\overline{p}{\rm He}^+$ two-photon resonance profiles,
we used the following nonlinear rate equations which describe a three-level model,
\begin{equation}
\label{sixfunc}
\begin{array}{@{}l}
\displaystyle
\frac{\partial\rho_{aa}}{\partial t}=-{\rm Im}(\Omega_{2M}\rho_{ab}),
\\[4mm]\displaystyle
\frac{\partial\rho_{bb}}{\partial t}={\rm Im}(\Omega_{2M}\rho_{ab})-{\rm Im}(\Omega_{1M}\rho_{bc}),
\\[4mm]\displaystyle
\frac{\partial\rho_{cc}}{\partial t}=-\gamma_c\rho_{cc}+{\rm Im}(\Omega_{1M}\rho_{bc}),
\\[4mm]\displaystyle
\frac{\partial\rho_{ab}}{\partial t}=-id_{ab}\rho_{ab}+i\frac{\Omega_{2M}}{2}(\rho_{aa}-\rho_{bb})+i\frac{\Omega_{1M}}{2}\rho_{ac},
\\[4mm]\displaystyle
\frac{\partial\rho_{bc}}{\partial t}=-id_{bc}\rho_{bc}+i\frac{\Omega_{1M}}{2}(\rho_{bb}-\rho_{cc})-i\frac{\Omega_{2M}}{2}\rho_{ac},
\\[4mm]\displaystyle
\frac{\partial\rho_{ac}}{\partial t}=-id_{ac}\rho_{ac}+i\frac{\Omega_{1M}}{2}\rho_{ab}-i\frac{\Omega_{2M}}{2}\rho_{bc}.
\end{array}
\end{equation}
Here the density matrix $\rho_{aa}$, $\rho_{bb}$, and $\rho_{cc}$ represent the
antiproton populations in the parent, intermediate, and daughter states. The mixing between pairs of states
induced by the lasers are denoted by $\rho_{ab}$, $\rho_{bc}$, and $\rho_{ac}$, the Auger
decay rate of the daughter state by $\gamma_c$. The three detunings that appear in Eq.~\ref{sixfunc}
can be calculated using the equations,
\begin{equation}
\begin{array}{@{}l}
\displaystyle
d_{ab}=E_b-E_a-\left(1+\frac{v_z}{c}\right)\omega_2-i\frac{\gamma_{a}+\gamma_{b}}{2},
\\[4mm]\displaystyle
d_{bc}=E_c-E_b-\left(1-\frac{v_z}{c}\right)\omega_1-i\frac{\gamma_{b}+\gamma_{c}}{2},
\\[4mm]\displaystyle
d_{ac}=E_c-E_a-\left(1+\frac{v_z}{c}\right)\omega_2-\left(1-\frac{v_z}{c}\right)\omega_1-i\frac{\gamma_{a}+\gamma_{c}}{2},
\end{array}
\end{equation}
where $v_z$ denotes the velocity component of the atom in the direction of the $\omega_2$-laser beam, $E_a$, $E_b$, and $E_c$ the binding energy of the
hyperfine states involved in the transition, and $\gamma_{a}$ and $\gamma_{b}$ the radiative rates of the
parent and intermediate states. The angular Rabi frequencies of single-photon transitions induced between the
parent and intermediate states of magnetic quantum number $M$, and the intermediate and daughter states
are respectively denoted by $\Omega_{2M}$ and $\Omega_{1M}$.

\begin{figure}[t]
\begin{center}
\vspace{-3mm}
\includegraphics[width=85mm]{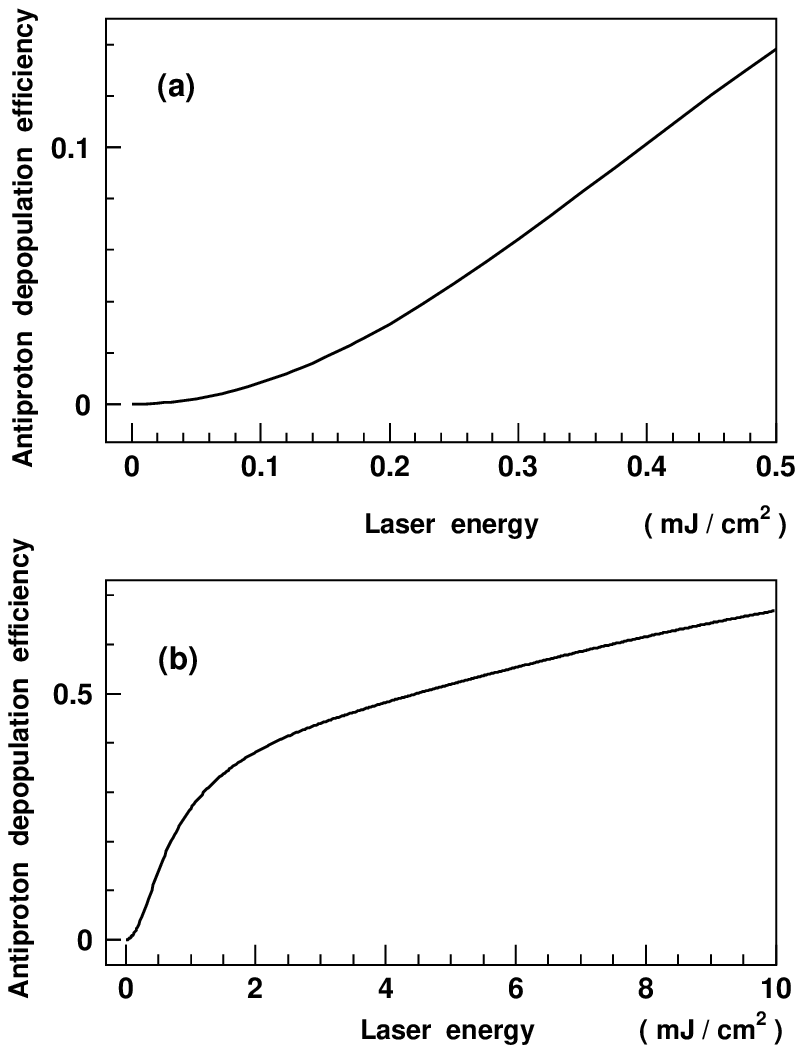}
\vspace{-3mm}
\end{center}
\caption{\label{simupower} Depopulation efficiency in the two-photon transition
$(n,L)=(36,34)$$\rightarrow$$(34,32)$ of $\overline{p}{\rm ^4He}^+$ at various
intensities of the two laser beams. The virtual intermediate state is
offset $\omega_d/2\pi=-12$ GHz away from state $(35,33)$,
so that the two laser frequencies coincide with the hyperfine component
$(S_{e},S_{\overline{p}})=(\uparrow\uparrow)$$\rightarrow$$(\uparrow\uparrow)$ of the resonance line.}
\end{figure}

Values of $\Omega_{2M}/2\pi$ (in s$^{-1}$) for the transition $(0,L,F,K,J,M)$$\rightarrow$$(q,L\!-\!1,F',K',J',M)$ in $\overline{p}{\rm ^3He}^+$ by linearly-polarized laser light of intensity $I_1$ (in W/cm$^2$) can be calculated as,
\begin{equation}
\begin{array}{@{}l}
\displaystyle
\frac{\Omega_{2M}}{2\pi} (\rm S.I.)=\sqrt{\frac{2I_1}{\epsilon_0c}}\frac{ea_0}{h}
\\[4mm]\displaystyle \hspace{6mm}
\times \Bigl|\langle 0LFKJM|\mathbf{d}|q(L\!-\!1)F'K'J'M\rangle\Bigr|\ (\rm a.u.)
\end{array}
\end{equation}
where the transition matrix element (in atomic units) can be derived using the Wigner $3j$- and $6j$-symbols,
\begin{equation}
\begin{array}{@{}l}
\displaystyle
\Bigl|\langle\, 0LFKJM|\,\mathbf{d}\,|q(L\!-\!1)F'K'J'M\,\rangle\Bigr|
\\[4mm]\displaystyle
=
\,\vrule width 0.7pt height 19pt depth 13pt
\langle\, 0L\|\mathbf{d}\|q(L\!-\!1)\,\rangle
\left(
   \begin{array}{@{\,}c@{\;\;}c@{\;\;}c@{\,}}
      J & 1 & J' \\
      M & 0 & -M
   \end{array}
\right)
\\[4mm]\displaystyle\hspace{10mm}
\times\sqrt{(2J\!+\!1)(2J'\!+\!1)}\>
\left\{
\begin{array}{@{\,}ccc@{\,}}
K' & J' & \frac{1}{2} \\
J & K & 1
\end{array}
\right\}
\\[4mm]\displaystyle\hspace{10mm}
\times\sqrt{(2K\!+\!1)(2K'\!+\!1)}
\left\{
\begin{array}{@{\,}ccc@{\,}}
F' & K' & \frac{1}{2} \\
K & F & 1
\end{array}
\right\}
\\[4mm]\displaystyle\hspace{10mm}
\times\sqrt{(2F\!+\!1)(2F'\!+\!1)}\>
\left\{
\begin{array}{@{\,}ccc@{\,}}
L' & F' & \frac{1}{2} \\
F & L & 1
\end{array}
\right\}
\,\vrule width 0.7pt height 19pt depth 13pt
\,.
\end{array}
\label{threesix}
\end{equation}
Eq.~(\ref{threesix}) only provides approximate values for the transition matrix elements since $F$ and $K$ are not exact quantum numbers of the three-body Hamiltonian. The results however agree with exact transition amplitudes within $1\%$, for all the $\overline{p}{\rm He}^+$ transitions of the type $(n,L,M)$$\rightarrow$$(n-1,L-1,M)$ studied here.

We numerically integrated Eq.~\ref{sixfunc} to simulate the antiprotons depopulated by the two laser beams from the resonance parent state to the daughter state via
a two-photon transition. Prior to laser irradiation, the antiprotons are assumed to uniformly populate the $\sim 70$ $M$-sublevels of the parent state.
The atoms follow a Maxwellian thermal distribution of temperature $T\!\sim\! 10$ K. The temporal profiles of the laser pulses are assumed to be roughly Gaussian with
pulse lengths $\Delta t\sim 100$ ns \cite{horiopt09}.

\begin{figure*}[htbp]
\begin{center}
\vspace{-3mm}
\includegraphics[width=160mm]{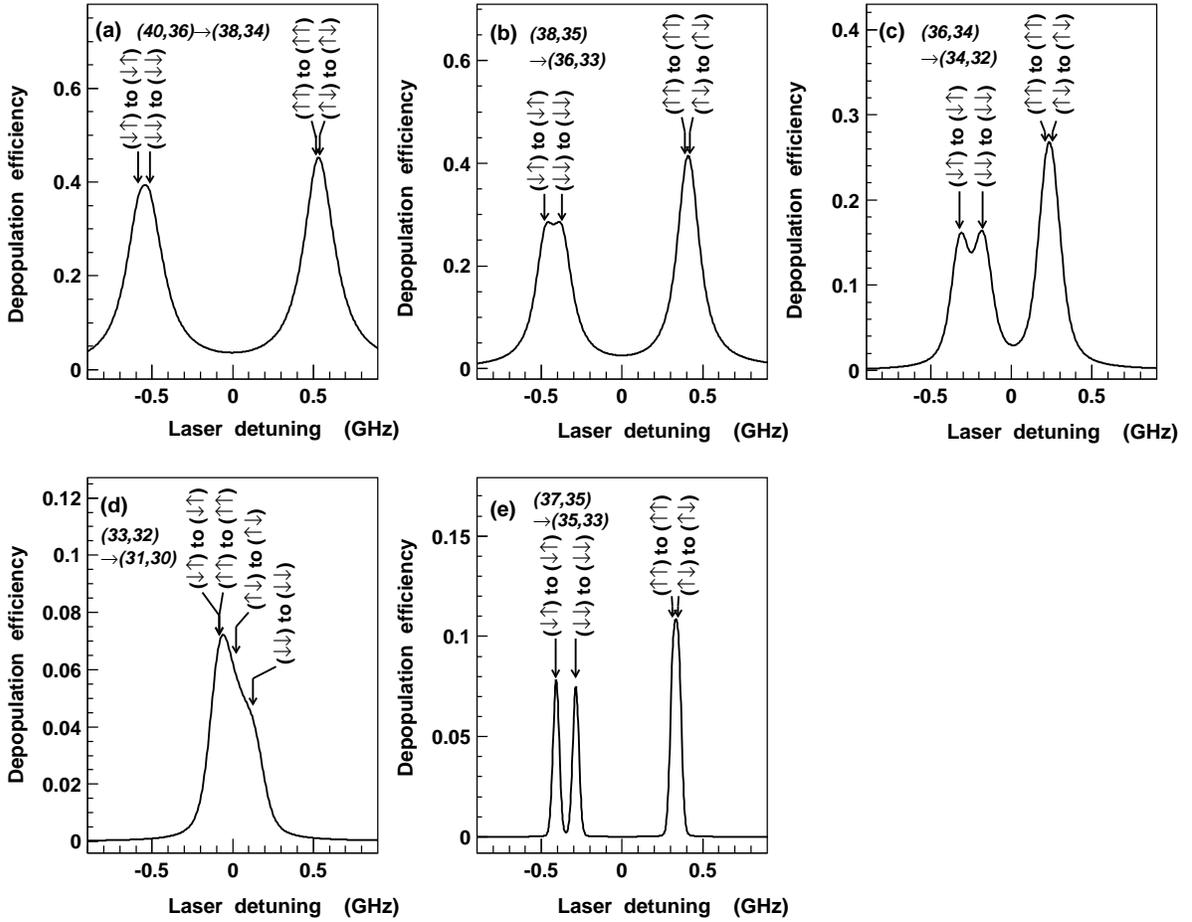}
\vspace{-9mm}
\end{center}
\caption{\label{simuprof} Simulated resonance profiles of five two-photon transitions in
$\overline{p}{\rm ^4He}^+$ excited at laser intensities $p\sim 1$ mJ/cm$^2$. The virtual
intermediate state was tuned $\Delta\omega_d/2\pi\sim -12$ GHz from a real state.
Four profiles (a)--(d) involve a resonance daughter state with lifetimes $\tau<10$ ns
against Auger decay. They are simulated assuming a temperature
$T\sim 10$ K of the atom.  The narrow resonance (e) involves a daughter state of much
longer ($\tau\sim 1$ $\mu{\rm s}$) lifetime, and was simulated at
$T\sim 1.5$ K. The positions of the four hyperfine lines are indicated by arrows,
together with the principal and angular momentum quantum numbers $(n,L)$ and
spin orientations $(S_{e},S_{\overline{p}})\rightarrow(S^{\prime}_{e},S^{\prime}_{\overline{p}})$
of the parent and daughter states.}
\end{figure*}

Fig.~\ref{simupower} (a)--(b) shows the efficiency $\varepsilon$ of the laser pulses depleting the population in
the parent state of the two-photon transition $(36,34)$$\rightarrow$$(34,32)$ of $\overline{p}{\rm ^4He}^+$
[i.e. $\varepsilon =1$ if the laser induces all the antiprotons occupying
state $(n,L)$ to annihilate, and $\varepsilon=0$ when no such annihilations occur].
The virtual intermediate state is offset $\omega_d/2\pi=-12$ GHz away from state $(35,33)$,
so that $\omega_1+\omega_2$ coincides with the hyperfine component
$(S_{e},S_{\overline{p}})=(\uparrow\uparrow)$$\rightarrow$$(\uparrow\uparrow)$ of the resonance line.
As the laser intensity is increased between $p=0$ and $0.2$ mJ/cm$^2$,
the $\varepsilon$-value increases quadratically as expected for a two-photon process.
It begins to saturate at $p>1$ mJ/cm$^2$ corresponding to $\varepsilon\sim 0.3$. Monte Carlo
simulations indicate that the two-photon resonance signal of $\varepsilon\sim 0.3$
would be strong enough for clear detection against the background caused by spontaneously annihilating antiprotons
\cite{horinim} with a signal-to-noise ratio of $>5$. Higher laser intensities would of course provide an even
stronger signal,  but power broadening effects then deteriorate the spectral resolution to several hundred MHz
and so this should be avoided for high-precision spectroscopy.

The resonance profiles of the two-photon transitions $(40,36)\rightarrow(38,34)$ of the $v=3$ cascade,
$(38,35)\rightarrow(36,33)$ of $v=2$, $(36,34)\rightarrow(34,32)$ of $v=1$, and
$(33,32)\rightarrow(31,30)$ of $v=0$ in $\overline{p}{\rm ^4He}^+$ at temperature
$T\sim 10$ K are shown in Figs.~\ref{simuprof} (a)--(d). These resonances have among the
largest transition amplitudes, and the Auger decay rates of the daughter states are large
$\gamma_A=(2.5-4)\times 10^{8}$ s$^{-1}$ which is a necessary condition to obtain a
strong annihilation signal \cite{horinim}. The intensities of the two lasers
are around $p\sim 1$ mJ/cm$^2$. The laser frequency $\omega_1$ is fixed to an offset
 $\Delta\omega_d/2\pi=-12$ GHz from the intermediate state, whereas
$\omega_2$ was scanned between -0.9 and 0.9 GHz around the two-photon resonance
defined by $\omega_1+\omega_2$.
In each simulated profile, the positions of the four hyperfine lines are indicated with
arrows together with the corresponding spin orientations $(S_{e},S_{\overline{p}})$.
The $\sim 200$ MHz linewidth of these profiles are primarily caused by the large
Auger width of the daughter states, and the residual Doppler and power broadening.

\begin{figure*}[htbp]
\begin{center}
\vspace{-3mm}
\includegraphics[width=160mm]{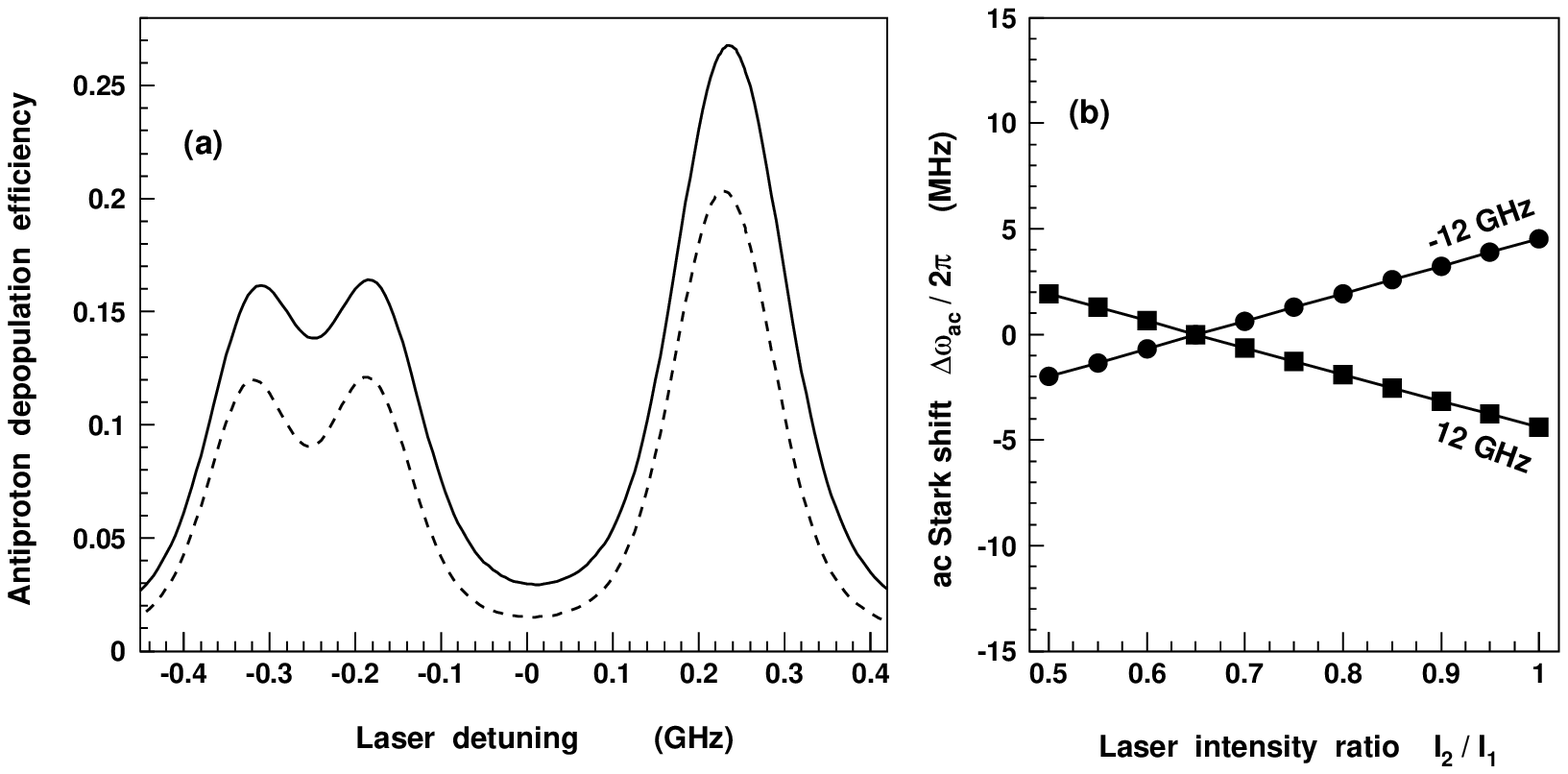}
\vspace{-5mm}
\end{center}
\caption{\label{simubloch} Simulated profiles of the resonance $(n,L)=(36,34)$$\rightarrow$$(34,32)$ in $\overline{p}{\rm ^4He}^+$,
with the frequency offset $\Delta\omega_d/2\pi\sim -12$ GHz and intensity ratio between the two
lasers $I_2/I_1=0.5$ (dashed lines) and 1 (solid lines) for constant $I_1 \sim 5\times 10^4$ W/cm$^2$ (a).
The atom is thermalized at $T\sim 10$ K.
The ac Stark shift in the simulated profiles for offsets $\Delta\omega_d/2\pi\sim -12$ GHz (solid circles) and $12$ GHz (squares),
as a function of $I_2/I_1$ (b).}
\end{figure*}

The resonance $(40,36)$$\rightarrow$$(38,34)$ shows a two-peak structure
[Fig.~\ref{simuprof}~(a)] with a frequency interval of $\sim 1.1$ GHz which arises from the
dominant spin-orbit interaction between $\mathbf{S}_{e}$ and $\mathbf{L}$.
Each peak is a superposition of two hyperfine lines with a few tens of MHz spacing
caused by a further interaction between the antiproton and electron spins. The asymmetric structure
of the profile of Fig.~\ref{simuprof}~(a) is due to the fact that the 25-MHz spacing between
the hyperfine lines $(S_{e},S_{\overline{p}})=(\uparrow\uparrow)$$\rightarrow$$(\uparrow\uparrow)$
and $(\uparrow\downarrow)$$\rightarrow$$(\uparrow\downarrow)$ are small compared to the
75-MHz spacing between $(\downarrow\uparrow)$$\rightarrow$$(\downarrow\uparrow)$ and
$(\downarrow\downarrow)$$\rightarrow$$(\downarrow\downarrow)$
The spacings between the hyperfine lines becomes gradually smaller for lower-lying
transitions involving states of smaller $n$- and $v$-values, e.g., $0.8$ and $0.5$ GHz for
$(38,35)$$\rightarrow$$(36,33)$ and $(36,34)$$\rightarrow$$(34,32)$. The hyperfine
lines can no longer be resolved for the lowest$-n$ transition $(33,32)$$\rightarrow$$(31,30)$
[Fig.~\ref{simubloch} (d)]. The low transition probability (Table~\ref{dipo}) of this resonance
causes the small depopulation efficiency $\varepsilon<0.1$ seen here; laser intensities
of $p\ge 2$ mJ/cm$^2$ would be needed to produce a sufficient experimental signal.

We expect the two UV transitions $(n,L)=(36,34)$$\rightarrow$$(34,32)$ and $(33,32)$$\rightarrow$$(31,30)$
in $\overline{p}{\rm ^4He}^+$ to yield the highest signal-to-noise ratios in laser spectroscopy
experiments. This is because the parent states $(36,34)$ and $(33,32)$ retain large antiproton
populations for long periods $t=3$--10 $\mu{\rm s}$ following $\overline{p}{\rm He}^+$ formation
\cite{mhori2002}. By comparison, cascade processes rapidly deplete the populations in higher
$n>37$ states within $1$-2 $\mu{\rm s}$, and so the associated two-photon spectroscopy signals
contain a large background due to the spontaneously annihilating $\overline{p}{\rm He}^+$ atoms
\cite{mhori2002,mhori2004}.

Higher experimental precisions on $\nu_{\rm exp}$ may be achieved by cooling the atoms to lower
temperature and by inducing two-photon transitions between pairs of $\overline{p}{\rm He}^+$
states with microsecond-scale lifetimes.
Fig.~\ref{simuprof} (e) shows the resonance $(37,35)$$\rightarrow$$(35,33)$ of
$\overline{p}{\rm ^4He}^+$ at temperature $T=1.5$ K. Both parent and daughter states
have lifetimes of $\tau\sim 1$ $\mu{\rm s}$, and so its natural linewidth $\sim 200$ kHz is two
orders of magnitude smaller than in the other resonances Figs.~\ref{simuprof} (a)--(d) studied here.
It is unfortunately difficult to measure this transition experimentally, as the present detection method
requires the daughter state to proceed rapidly to antiproton annihilation. Cooling the $\overline{p}{\rm He}^+$
atoms to such low temperatures may be technically challenging.

\begin{figure*}[htbp]
\begin{center}
\vspace{-5mm}
\includegraphics[width=160mm]{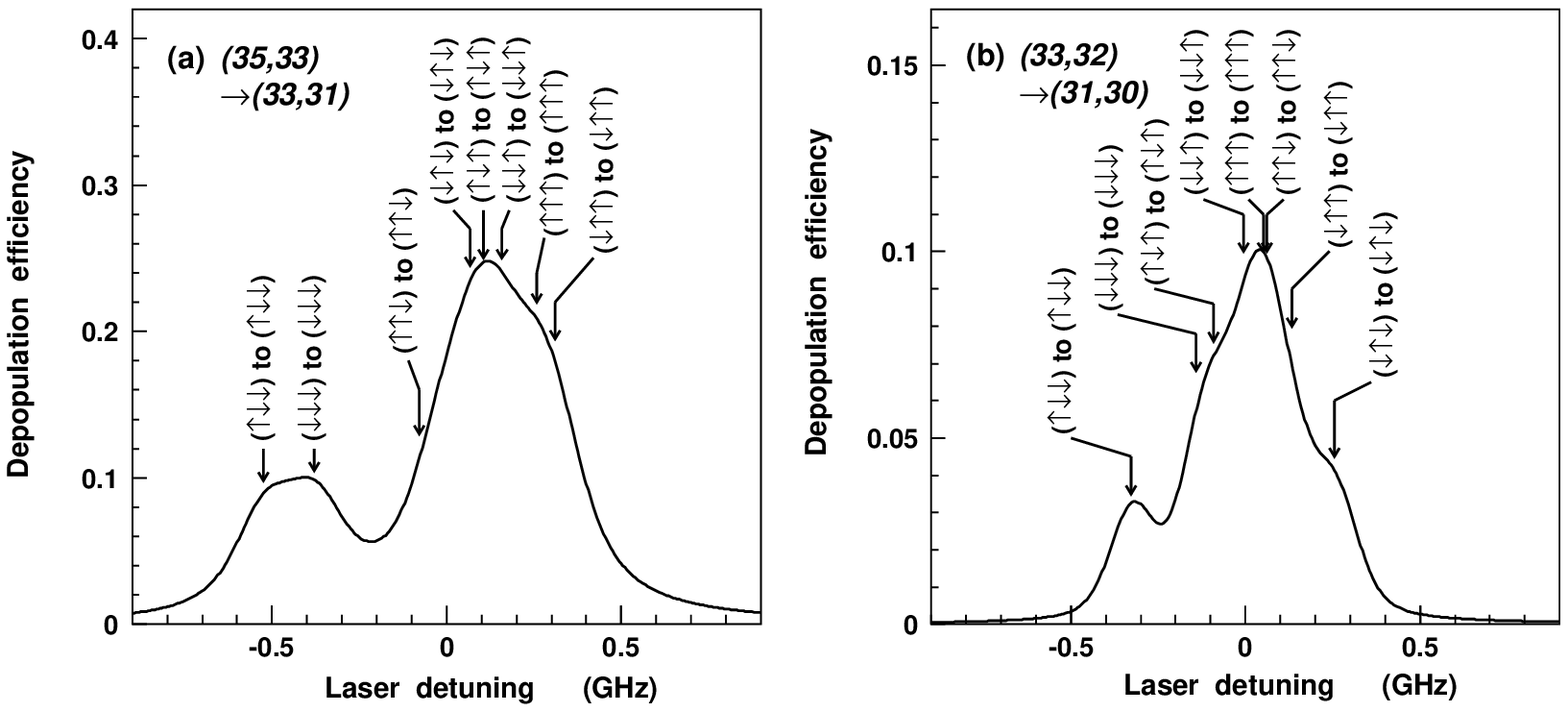}
\vspace{-5mm}
\end{center}
\caption{\label{simuprof3}
Simulated resonance profiles of two two-photon transitions in
$\overline{p}{\rm ^3He}^+$ excited at laser intensities $p\sim 1$ mJ/cm$^2$ and
temperature $T\sim 10$ K. The virtual
intermediate state was tuned $\Delta\omega_d/2\pi\sim -12$ GHz from a real state.
The positions of the eight hyperfine lines are indicated by arrows, together with the
principal and angular momentum quantum numbers $(n,L)$ and
spin orientations $(S_{e},S_{h}, S_{\overline{p}})$$\rightarrow$$(S^{\prime}_{e},S^{\prime}_{h},S^{\prime}_{\overline{p}})$
of the resonance parent and daughter states.
}
\end{figure*}

The profiles of two $\overline{p}{\rm ^3He}^+$ resonances
which are expected to yield the highest signal to noise ratios \cite{mhori2002}
$(35,33)$$\rightarrow$$(33,31)$ and $(33,32)$$\rightarrow$$(31,30)$ at
temperature $T\sim 10$ K are shown in Figs.~\ref{simuprof3} (a)--(b).
The positions of the eight hyperfine lines and their spin configurations
$(S_{e},S_{h},S_{\overline{p}})$ are indicated by arrows. Due to the
large number of partially overlapping lines, it may be difficult to determine
the $\nu_{\rm exp}$-values for $\overline{p}{\rm ^3He}^+$ with a similar
level of precision as in $\overline{p}{\rm ^4He}^+$. The problem would be
especially acute in the case of $(33,32)$$\rightarrow$$(31,30)$ [Fig.~\ref{simuprof3} (b)]
which contain 8 sublines within a relatively small 0.6-GHz interval.

We finally use these numerical simulations to determine the ac Stark shift under
realistic experimental conditions. Fig.~\ref{simubloch} (a) shows the profiles of the
resonance $(36,34)$$\rightarrow$$(34,32)$ of $\overline{p}{\rm ^4He}^+$ at
temperature $T\!\sim\! 10$ K and laser offsets $\Delta\omega_d/2\pi\!=\!-12$ GHz.
They were calculated at two combinations of the laser
intensities: $I_1=5\!\times\!10^4$ W/cm$^2$ and $I_2=2.5\!\times\!10^4$ W/cm$^2$
(broken lines) and $I_1\!=\!I_2\!=\!5\!\times\! 10^4$ W/cm$^2$
(solid lines). As $I_2/I_1$ is increased, the transition frequency shifts to larger values.
In Fig.~\ref{simubloch} (b), the ac Stark shifts $\Delta\omega_{\rm ac}/2\pi$ determined
from the simulated profiles of Fig.~\ref{simubloch} (a) at laser offsets $\Delta\omega_d/2\pi=-12$ GHz
are plotted using filled circles. It increases linearly from $-2$ MHz at $I_2/I_1=0.5$, to 5 MHz at $I_2/I_1=1$.
A similar plot for offset $\Delta\omega_d/2\pi=12$ GHz is shown using filled squares.
The two calculated sets of ac Stark shifts are of equal magnitude and opposite sign, the minimum
occurring around $I_2/I_1\sim 0.65$.

\section{Conclusions}

We conclude that two-photon transitions in $\overline{p}{\rm He}^+$ of the
type $(n,L)$$\rightarrow$$(n\!-\!2,L\!-\!2)$ can indeed be induced using two counterpropagating
nanosecond laser pulses of intensity $\sim 1$ mJ/cm$^2$, for cases where the virtual
intermediate state is tuned within $\left|\Delta\omega_d/2\pi\right|=10$--20 GHz of the real state $(n\!-\!1,L\!-\!1)$.
The spectral resolution of the measured resonances should increase by an order of
magnitude or more compared to conventional single-photon spectroscopy.
The ac Stark shifts at these experimental conditions can reach several MHz or more,
but this can be minimized by carefully adjusting the relative intensities of the two
laser beams. Any remaining shift can be canceled by comparing the resonance profiles
measured at positive and negative offsets $\pm\Delta\omega_d$ of the virtual intermediate
state from the real state. In practice, this can be done by, e.g., using a frequency comb
\cite{udem} to accurately control the frequencies $\omega_1$ and $\omega_2$ of the
counterpropagating laser beams. The UV two-photon transitions $(36,34)$$\rightarrow$$(34,32)$
and $(33,32)$$\rightarrow$$(31,30)$ in $\overline{p}{\rm ^4He}^+$, and
$(35,33)$$\rightarrow$$(33,31)$ in $\overline{p}{\rm ^3He}^+$ are expected to yield
particularly strong resonance signals that can be precisely measured.

\acknowledgements
We are indebted to R.S.~Hayano. This work was supported by the European Science Foundation and the Deutsche Forschungsgemeinschaft (DFG), the Munich Advanced Photonics (MAP) cluster of DFG, the Research Grants in the Natural Sciences of the Mitsubishi Foundation, and the Initiative Grant No.~08-02-00341 of the Russian Foundation for Basic Research.

\end{document}